\documentclass{ws-procs9x6}
\usepackage{hyperref}
\usepackage{epsfig}

\begin{document}

\title{Innovative Experimental Particle Physics through Technological Advances
--\\
Past, Present and Future}

\author{H.~W.~K.~Cheung}

\address{Fermi National Accelerator Laboratory, \\
P.O. Box 500, \\ 
Batavia, IL 60510-0500, USA\\ 
E-mail: cheung@fnal.gov}

\maketitle

\abstracts{
This mini-course gives an introduction to the techniques used in 
experimental particle physics with an emphasis on the impact of
technological advances. The basic detector types and particle accelerator
facilities will be briefly covered with examples of their use and
with comparisons. The mini-course ends with what can be expected in the
near future from current technology advances. The mini-course is intended
for graduate students and post-docs and as an introduction to 
experimental techniques for theorists.}

\section{Introduction}\label{sec:intro}
Despite the fancy title of this mini-course, the intention is to give
a brief introduction to experimental particle physics. 
Since there are already some excellent introductions to this
topic and some textbooks that cover various detectors in detail,
a more informal approach to the topic is given in this mini-course.
Some basic detector elements are covered while reviewing examples of real
experiments, and experimental techniques are introduced by comparing
competing experiments. Some aspects of
experimental design are
also briefly reviewed. Hopefully this will provide a more engaging 
introduction to the subject than a traditional textbook. This
short mini-course cannot replace a real experimental physics course;
the reader is just given a taste. Unfortunately the lack of space for
this writeup means that not even the basic detection methods and
detector types can be described. Instead detector types in {\em italics} will
be briefly described in a glossary 
at the end of this writeup. For further reading,
the reader can find
the relevant physics of particle
interactions and detailed descriptions of many different types of
particle detectors in a number of textbooks
and articles\rlap.\,\cite{ref:textbooks}

Although this mini-course is devoted to the physics impact of
some significant technological advances, it should be noted that
the improvements in experimental techniques usually progress in steady
steps. Quite often advances are linked to steady progress in the following
areas:

\begin{itemize}
\item Higher energy available or/and higher production rate.

\item Improvements in momentum or/and position resolution.

\item Better particle identification methods.

\item Increase in detector coverage or energy resolution.

\item More powerful signal extraction from background.

\item Higher accuracy (due to increase in data statistics, reduction of 
experimental systematic 
uncertainties, or reduction in theoretical uncertainties). 
\end{itemize}

\begin{figure}[ht] 
\centerline{\epsfxsize=4.1in\epsfbox{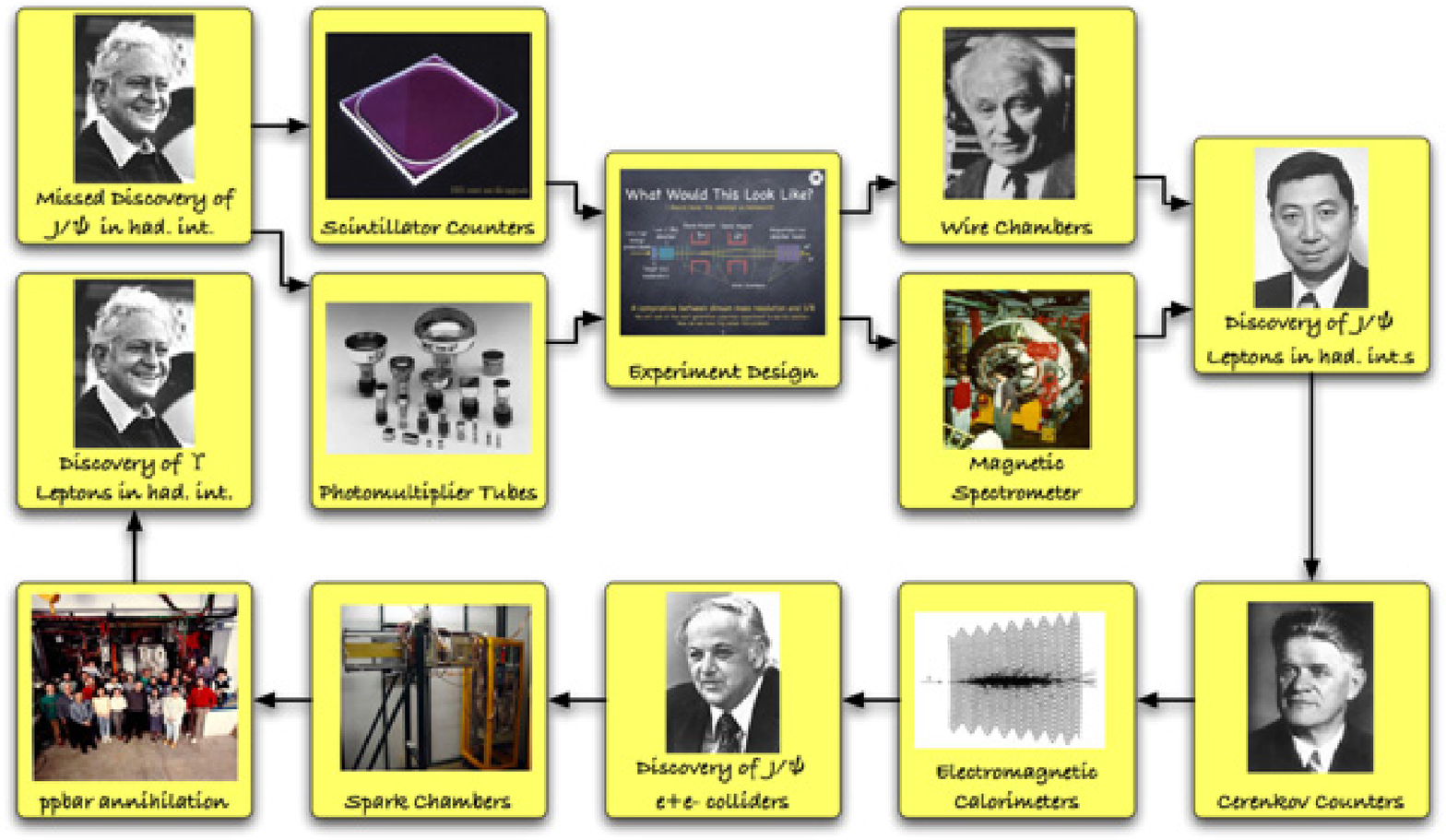}}
\centerline{\epsfxsize=4.1in\epsfbox{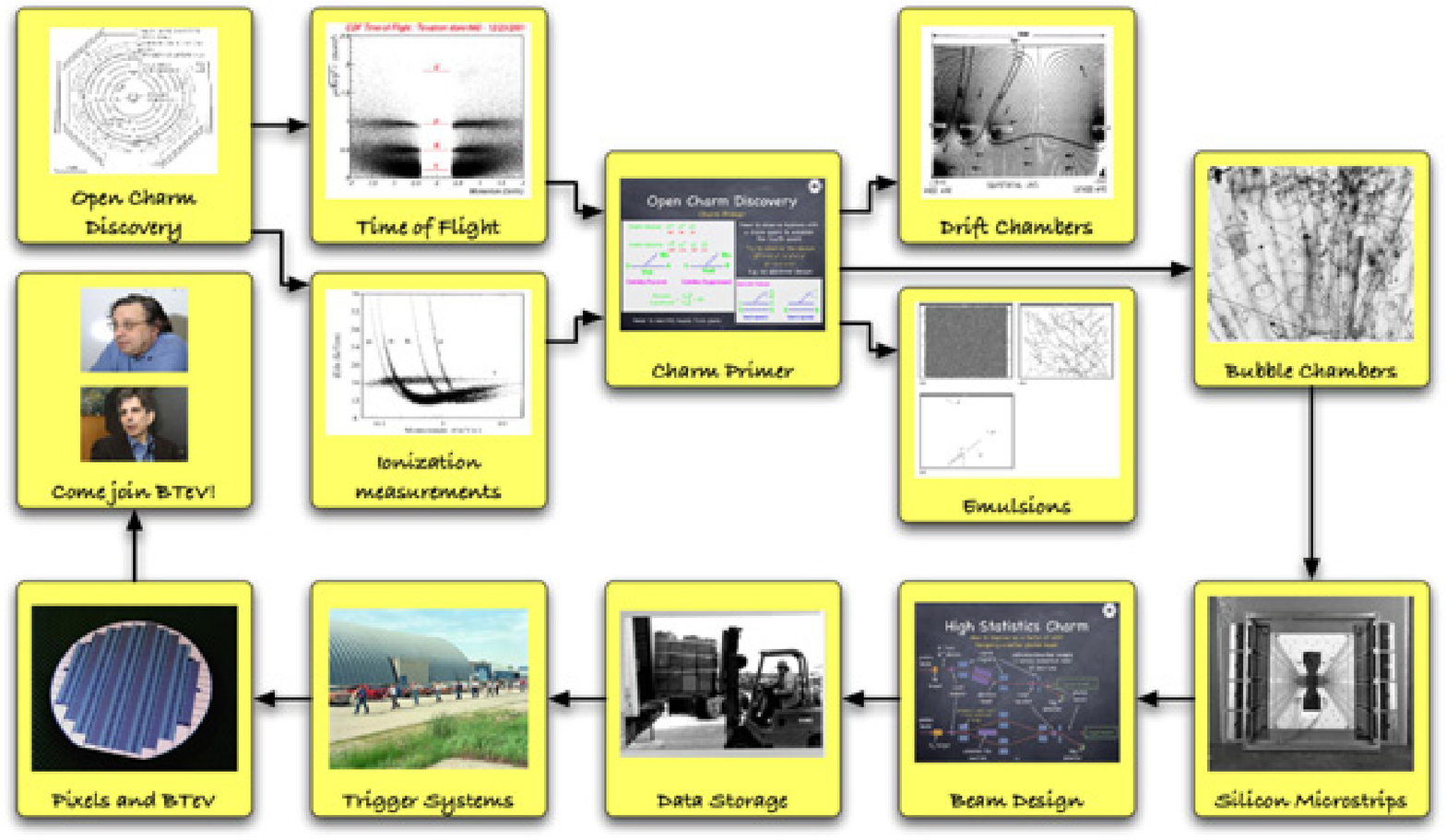}}    
\caption{Outline for the the mini-course: (top) lecture 1; (bottom)
lecture 2. \label{fg:outline}}
\end{figure}

Discoveries are often made through a series of incremental steps,
though of course the discoveries themselves can be in a surprising
direction! The topics I have chosen for the
two lectures of this mini-course is the
discovery and subsequent study of the charm quark, 
and the future of bottom quark physics. The outlines for the two lectures are
illustrated in Fig.~\ref{fg:outline}.

\section{Part I: Discovery of Charm}\label{sec:charmdiscovery}

The discovery of the $J/\psi$ meson is well documented by many books and
articles\cite{ref:charmdiscovery} 
as well as in the Nobel lectures of Ting\cite{ref:tingnobel}
and Richter\rlap.\,\cite{ref:richternobel} Besides being a great classic story
of discovery, we can also use it to illustrate some of the detection
techniques and the physics and
ideas behind the design of the experiments involved.

\subsection{A Missed Opportunity: Resolutions Matter!}\label{sec:resolutions1}

Since the leptons, electrons and muons, are basically point-like,
stable or long-lived, and
interact primarily via the well understood and calculable electroweak
force, they have served as the ``eyes'' in probing many experimental
processes. One of those processes under study in the 1970's was hadron
interactions. The interests in this study included 
the investigation of the electromagnetic
structure of hadrons, the study 
of the then-called ``Heavy photons'' $\rho$, $\omega$ and $\phi$
and the search for additional ones, as well as the search
for the neutral intermediate vector boson, the $Z^0$.

One experiment doing such a study offers a lesson on the importance of
experimental resolution. This was an experiment
at Brookhaven
National Laboratory (BNL) using the Alternating-Gradient
Synchrotron (AGS) carried out by Leon Lederman's group. They performed
studies of $p+U\rightarrow \mu^+\mu^-X$ and missed discovering
the $J/\psi$ in 1970, four years before the actual discovery.

\begin{figure}[ht] 
\centerline{\epsfxsize=4.1in\epsfbox{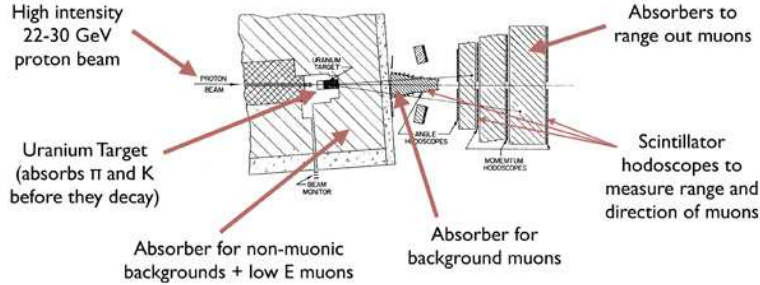}}    
\caption{Diagram of the spectrometer for the Lederman 1970 
experiment\rlap.\,\protect\cite{ref:lederman1970} \label{fg:lederman1970a}}
\end{figure}

A diagram showing Lederman's 1970 experiment is given in 
Fig.~\ref{fg:lederman1970a}\rlap.\,\cite{ref:lederman1970,ref:ledermannobel}
The experiment was to study the interaction
of the 22-30~GeV proton beam on a Uranium target. The aim was to detect
a pair of oppositely charged muons coming from the interaction. 

The emphasis of this experiment was to get a clean signature for muons
directly produced in the target. The main background to eliminate was
muons from the decay of pions and kaons. A high atomic number target
like Uranium has a short interaction length which serves to both
cause a lot of the proton beam to interact and also to absorb
pions and kaons produced in the interactions before they can decay.
This is followed by additional material to absorb non-muonic backgrounds and
low energy muons. Muons from hadron decay typically have lower energy 
than those directly produced in the primary proton-Uranium interaction.
Another specially shaped heavy absorber serves to absorb more
background muons while 
{\em scintillator hodoscopes} measure the
direction of the surviving muons. The final material at the end of the
detector serves to measure the range and therefore the energy of the
muons. 

Although all the absorber material helps to give a much cleaner
sample of dimuon events, it also causes a lot of multiple Coulomb
scattering (MCS), especially as the material is of high $Z$ and therefore has
short radiation length. This large MCS limited the dimuon mass resolution
at 3~GeV/c$^2$ to about $\approx$13\%, or $\approx$400~MeV/c$^2$.
Even with all the absorber the signal-to-background (S/B) is relatively small.
The S/B at low dimuon mass ($\approx$2~GeV/c$^2$)
was about 2\%, increasing up to 50\%\ at higher dimuon 
mass ($\approx$5~GeV/c$^2$).
This means relatively large background subtractions are needed. Another
concern was the low acceptance and efficiency at low dimuon mass which
therefore needed larger corrections. 

\begin{figure}[ht] 
\centerline{\epsfxsize=3.5in\epsfbox{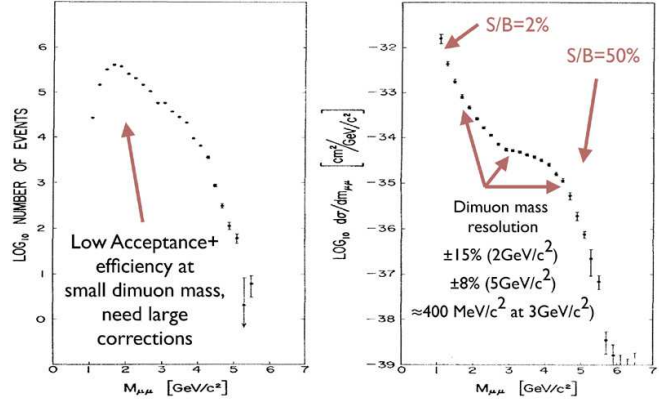}}    
\caption{Results for the dimuon mass spectrum from the Lederman 1970 
experiment\rlap.\,\protect\cite{ref:lederman1970}
(left) raw distribution; (right) corrected spectrum.
\label{fg:ledermandimumass}}
\end{figure}

The raw and corrected dimuon mass distributions are given in 
Fig.~\ref{fg:ledermandimumass}. Although the large background
subtraction and the uncertainty in the correction might have
contributed to the missed discovery of a peak at the $J/\psi$ mass, if
the dimuon mass resolution were sufficiently better, the $J/\psi$ peak
would still have been observed. The experimenters did many tests and
gave limits for a narrow state, but they had to conclude in the end that
there was ``no forcing evidence of resonant structure.''

\subsection{Elements of Experimental Design}\label{sec:expdesign}

With hindsight how would we change Lederman's 1970 experiment so we
could observe the $J/\psi$? Instead of leaving it as a task to the reader,
it is instructive to go through this in a little detail. The most obvious
things to include in a redesign are the following:

\begin{romanlist}
\item Improve the momentum resolution, which means having less material
for MCS, using a magnet for momentum determination and 
using a finer spatial resolution
detector than a scintillator hodoscope.
\item Increase the S/B, which means separating muons
better from hadrons and enriching the sample of dimuons {\em vs.} single
muons.
\item Achieve better acceptance and efficiency, which for a study of
the dimuon mass spectrum means obtaining a flatter efficiency as a
function of dimuon mass. A smooth efficiency across the dimuon mass is
probably fine as long as the efficiency (correction) is well understood.
\end{romanlist}

The average angular deflection due to
MCS of directly produced (signal) muons is given by
$\theta_{\mathrm{MCS}}\sim 
(Z_{\mathrm{Target}}/p_{\mu})\sqrt{L_{\mathrm{Target}}}
\sim (1/p_{\mu})\times (L_{\mathrm{Target}}/\lambda_0)$. 
Where $\lambda_0$ is the radiation length. So to reduce the
effects of MCS one should select a short target with long radiation length
and use as high a beam energy as possible to produce more 
higher momentum signal muons. Targets with low $Z$/$A$ will have longer
radiation lengths but they also have lower density and thus a longer
target would be needed to get the same number of 
inelastic proton interactions
in the target. A large signal sample needs a target
with high atomic number, 
since the dimuon signal rate $\sim A_{\mathrm{Target}}$. Another
consideration is that the
absorption probability for pions and kaons $\sim A_{\mathrm{Target}}^{0.7}$.
Thus the S/B would increase with heavier targets and dense targets.
One would need to do a Monte Carlo simulation to study what target
material is optimal. 

If the effects of MCS can be sufficiently reduced
we would need to determine the momentum of the muon more precisely.
This can be achieved with a
{\em magnetic spectrometer} where the deflection
in a known magnetic field can give the magnitude of 
the momentum. The angle of deflection can be obtained with low mass 
{\em Multiwire Proportional Chambers} ({\em MWPC's})
placed before any of the hadron absorbers.

\begin{figure}[ht] 
\centerline{\epsfxsize=4.1in\epsfbox{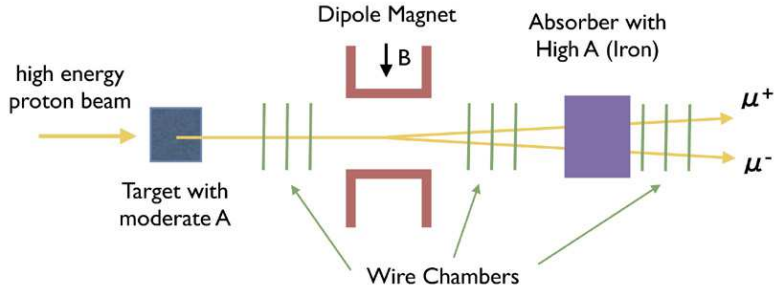}}    
\caption{Schematic for one possible redesign of a detector studying dimuons
where the main consideration was with improving
the dimuon mass resolution. \label{fg:redesign1}}
\end{figure}

An initial redesign of the
detector might look 
something like the schematic in Fig.~\ref{fg:redesign1}.
However this does not have all the absorbing material of the 1970
Lederman detector to reduce background muons. For that experiment the
S/B$\sim 0.04$, while in our initial design it could be as small
as 10$^{-6}$! To see how the S/B could be improved one has to
consider the sources of background. The main ones are given below:

\begin{romanlist}
\item Direct single muons -- these should be relatively small at
AGS energies since the production would be through electroweak
processes. Production via the decay of $\tau$ leptons or charm particles
is of the same level as the $J/\psi$, so getting an
accidental dimuon pair through
these decays should cause a negligible background.

\item Muons from decays of hadrons -- these happen early due to an
exponential decay and should therefore be absorbed early before they
can decay. Also lower momentum hadrons will decay relatively sooner
and thus make up a larger fraction of the decay muon background. One
could try to reject softer muons from the data analysis. Also one could make
multiple measurements of the momentum to reject muons from decay
in flight.

\item Hadrons from ``punch through'' -- a signal in a
detector element placed after an absorber can arise due to the end
of a hadronic shower leaking through
the absorber. One can try to detect this by
having multi-absorber/detection layers which can be used to recognize
a hadronic shower signal from a typically minimum ionizing muon
signature. One can also try to momentum analyze through a magnetic
absorber.
\end{romanlist}

\begin{figure}[ht] 
\centerline{\epsfxsize=4.1in\epsfbox{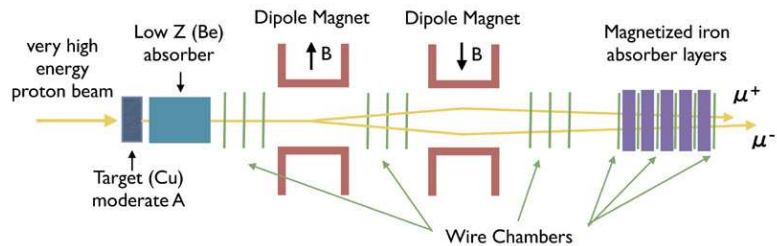}}    
\caption{Schematic for a second
possible redesign of a detector studying dimuons
where S/B was included as a consideration as well as
the dimuon mass resolution. \label{fg:redesign2}}
\end{figure}

Another example of a revised design for the dimuon detector, this time
also taking into consideration the S/B
is shown in Fig.~\ref{fg:redesign2}. It is seen that
to do better than the 1970 Lederman experiment one needs a more
complicated detector with considerably more advanced detectors. Even so
one can see there is still a compromise made between getting the best
S/B and the best dimuon mass resolution. One would need to do a serious
Monte Carlo simulation to determine the optimal choices.

We have only really touched on the elements of experimental design.
For example timing considerations have been completely ignored
and we have assumed the wire chambers can handle the necessary rates.
Instead of pursuing this further, and also before showing you Lederman's
solution, we first turn to see how Ting solves this
experimental design problem.

\subsection{Ting's Solution}

It was recognized that the same physics could be studied by
observing pairs
of electrons instead of dimuons. Electrons can be produced by
the same decays and have the
same $J^{PC}$ as muons, thus dielectrons should
also be produced by the $J^P=1^{-}$ $\rho$, $\omega$, $\phi$ and
$J/\psi$. However electrons differ in that they are about 200 times lighter
than muons. This greatly changes the considerations for a detector
designed to measure dielectron pairs compared to dimuon pairs.
Although kaon and pion decays are no longer a serious source
of background for a study of dielectrons, the
electrons undergo much more scattering and absorption than muons. Thus
the choice of materials and the detector types used to identify and
track electrons is quite different.

\begin{figure}[ht] 
\centerline{\epsfxsize=4.1in\epsfbox{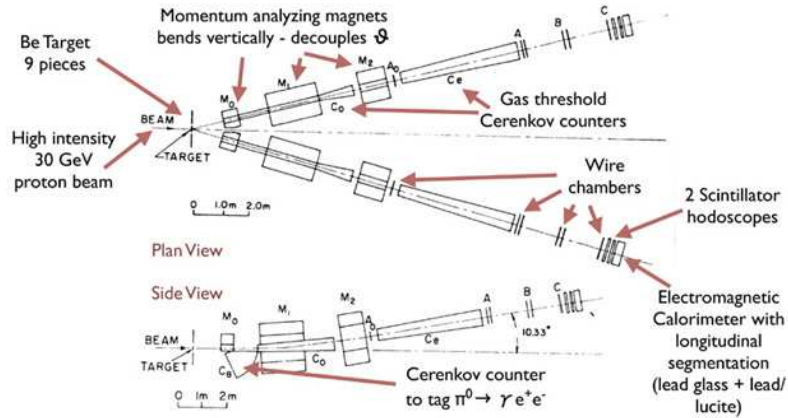}}    
\caption{Schematic of the detector for Ting's $J/\psi$ observation
experiment\rlap.\,\protect\cite{ref:tingnobel} \label{fg:tingspect}}
\end{figure}

Figure \ref{fg:tingspect} shows a schematic of the spectrometer used
by Ting in his $J/\psi$ observation experiment. It can be seen that
the spectrometer is quite complex. Low $Z$ beryllium targets 
and low mass {\em MWPC's} are used
to avoid too many photons converting to $e^+e^-$ pairs. The use of a
{\em multi-magnet spectrometer} and 
{\em MWPC's} helped to achieve a very fine
mass resolution of $\approx 5$~MeV/c$^2$. Without electron identification
the S/B would have been $\sim 10^{-6}$, thus a relative background rejection
of $10^6$--$10^8$ was needed. Typically a single particle identification
detector can achieve a relative background rejection of $10^2$--$10^3$
so multiple systems were combined. Both {\em \v Cerenkov counters} and
{\em electromagnetic calorimeters} were used to identify electons.
A special {\em \v Cerenkov counter} was used to specifically 
reject background from $\pi^0\rightarrow\gamma e^+e^-$. A relative
background rejection of $10^8$ was achieved and, together with a fine
dielectron mass resolution, a spectacularly narrow and clean $J/\psi$
signal was seen. The results and details of how this analysis was done
are well documented in Ting's Nobel lecture\cite{ref:tingnobel}
and the published papers\rlap.\,\cite{ref:tingpapers}

\subsection{Richter's Solution}

There is another half of the $J/\psi$ discovery story that cannot be
covered in this writeup because of insufficient space. Revealed in that
half would be additional important experimental
techniques. For example Richter's
observation of the $J/\psi$ was made in an $e^+e^-$ collider, a relatively
new innovation at that time. A nearly $4\pi$ detector was used
including {\em wire spark chambers} and 
{\em electromagnetic shower counters}. The $J/\psi$ mass resolution was much
better since it was governed
by knowledge of the beam energy and thus the widths of states can be
much better measured. That half of the story and the subsequent studies
are well documented in Richter's Nobel lecture\rlap.\,\cite{ref:richternobel}

\subsection{Improving Charmonium Spectroscopy}

\begin{figure}[ht] 
\centerline{\epsfig{file=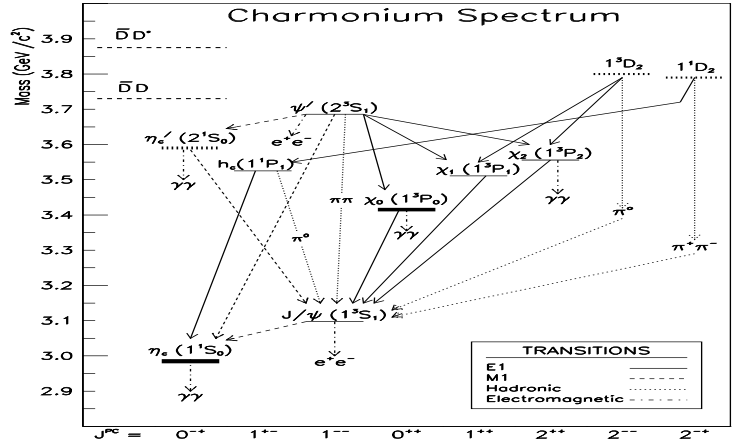,width=4.0in,height=2.4in}}
\caption{Schematic of the
charmonium energy levels. \label{fg:charmoniumlevels}}
\end{figure}

An $e^+e^-$ collider is an excellent study tool. This was recognized by
Ting as well as by Richter. It is specially well suited to perform
detailed studies of vector particles once their mass is known. This has
been the case for charmonium, for bottomonium and for the $Z^0$.
Narrow states with unknown masses are difficult to find. However
special modifications were made to
the SPEAR $e^+e^-$ storage ring to enable scans in energy in a relatively
short time. This enabled the discovery of the $J/\psi$ by Richter's team
as well as some of the charmonium excited
states. A disadvantage is that the $e^+e^-$ collisions can only directly
produce states with $J^P=1^-$, thus only these are measured with
fine resolution. While some of the
non-($J^P=1^-$) charmonium states
could be observed through the decays of the $\psi^{\prime}$,
see Fig.~\ref{fg:charmoniumlevels}, the measurements of their masses
and widths can no longer be obtained with just the knowledge
of the beam energies. 

Further improvement in the knowledge of the charmonium spectrum has
been achieved by using low energy $p\bar{p}$ collisions in an
antiproton accumulator. The first of these experiments was done at
the CERN ISR in R704, then in E760 and E835 at the Fermilab
antiproton accumulator\rlap.\,\cite{ref:e835} 
In the Fermilab experiments, a hydrogen
gas-jet target is used and the antiproton beam is tuned to produce
and precisely measure charmonium states of any $J^P$. The
charmonium states are tagged by their electromagnetic decays
using {\em lead glass shower counters} and {\em scintillating fibres}.
All that is needed is to 
recognize signal from background. The actual mass and width
measurements is determined with exquisite ($\sim 0.01$\%)
resolution due to excellent
knowledge of the beam energy.

\subsection{Lederman's Two Solutions}

We conclude the first part of the mini-course with two of Lederman's
solutions to the dilepton experiment design problem. The first is his
1976 experiment that looked at $e^+e^-$ pairs using a higher
energy beam running at Fermilab\rlap.\,\cite{ref:lederman1976}
This detector was a relatively simple experiment using a {\em magnetic
spectrometer} for momentum determination and a 
{\em lead-glass calorimeter} for electron identification.
Although the $J/\psi$ was clearly visible in this experiment, the
background was still relatively high. A cluster of events was observed
at $M_{e^+e^-}\approx 6$~GeV/c$^2$ which lead to a claim of a possible
observation of a narrow peak at this mass. What was observed was most
likely a background fluctuation\rlap.\,\cite{ref:ledermannobel}
Lederman's second solution was
a 1977 experiment to look at dimuons using a far more complicated
detector and again running at Fermilab\rlap.\,\cite{ref:lederman1977}
This was a far more successful experiment in which the
first observation of the $\Upsilon$
was made, the first indication of a new fifth
quark. Unfortunately observations of new quarks were 
apparently no longer deemed
worthy of a Nobel prize by this time. However the reader need not feel
too bad for Leon Lederman since he was awarded the Nobel prize anyway in
1988, sharing it with Melvin Schwartz and Jack Steinberger
for their use of neutrino beams and
discovering a second type of neutrino, the muon 
neutrino\rlap.\,\cite{ref:ledermanneutrino}


\section{Part II: More on Charm and Bottom Quarks}

In Part I the discovery of charm was used to introduce some basic
detectors components. 
{\em Scintillators}, {\em Photomultiplier Tubes},
{\em wire chambers}, 
{\em magnetic spectrometers}, {\em \v Cerenkov counters},
and 
{\em electromagnetic calorimeters} were mentioned. In
Part II additional experimental topics are covered, namely 
the following: particle
identification systems; the use of precision position
detectors to observe detached vertices; the use of different
beam types; and the evolution of trigger systems. The story for this
part of the mini-course is the advancement of detection of particles
containing charm and bottom quarks.
The outline of this part is illustrated in the
bottom section of Fig.~\ref{fg:outline}.

\subsection{Open Charm Discovery}\label{sec:opencharm}

In Sec.~\ref{sec:charmdiscovery}\ we introduced the discovery of the
charmonium ($c\bar{c}$) states where the charm quantum number is hidden.
The charm quark explanation of the observed narrow states became
universally accepted once states with open charm were discovered.

The two most commonly produced charm mesons are the $D^0$ ($c\bar{u}$)
and the $D^+$ ($c\bar{d}$). The charm quark decays quickly to either
a strange quark or a down quark. The ratio of the rates for these
two decays is given by the ratio of the
square of two CKM matrix elements:
$\Gamma(c\rightarrow sW^{\ast})/\Gamma(c\rightarrow dW^{\ast})\sim
|V_{cs}|^2/|V_{cd}|^2\approx 20$.
The $c\rightarrow sW^{\ast}$ decay is called Cabibbo favoured while the
$c\rightarrow dW^{\ast}$ is Cabibbo suppressed. The virtual $W$ can
decay to either quarks or leptons. Thus most of the $D^0$ and $D^+$ mesons
decay to states with a strange quark. The easiest decay modes to 
reconstruct are the all charged modes: 
$D^+\rightarrow K^-\pi^+\pi^+$;
$D^0\rightarrow K^-\pi^+$; and $D^0\rightarrow K^-\pi^+\pi^+\pi^-$.
Since pions are the more copiously produced hadrons in an interaction,
one needs to distinguish kaons from pions to observe these open charm
signals.

Besides {\em \v Cerenkov counters} there are other particle identification
methods for charged hadrons. One example is a {\em Time-of-Flight} ({\em TOF})
detector, this was used in the discovery of open charm
two years after the discovery of the $J/\psi$. The discovery was made using
the Mark I experiment at the SPEAR $e^+e^-$ collider, the same 
spectrometer which was used in the
discovery of the $J/\psi$. The $e^+e^-$ collider gives an inherently
lower background than hadron-hadron collisions since the electron and
positron annihilate completely. 
However the $D^+\rightarrow K^-\pi^+\pi^+$ and
$D^0\rightarrow K^-\pi^+$, $K^-\pi^+\pi^+\pi^-$ were only discovered after
the collection of additional data and using the {\em TOF} system to
separate kaons from pions.

A {\em TOF} system works by measuring the time it takes for a charged
particle to travel between two points. For particles of the same
momentum, this time difference depends on the particle's mass.

\begin{figure}[ht] 
\centerline{\epsfxsize=1.8in\epsfbox{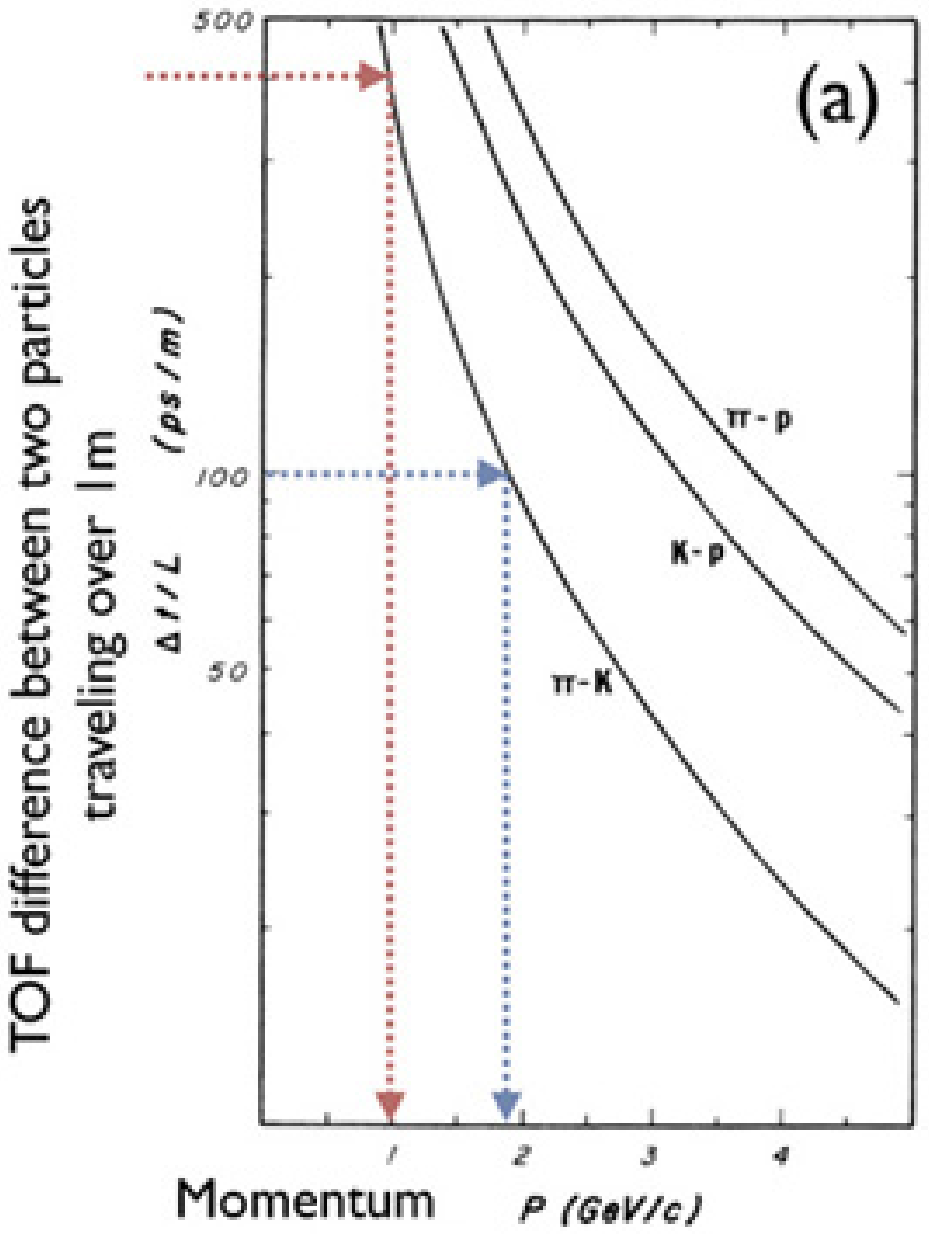}
\epsfxsize=2.1in\epsfbox{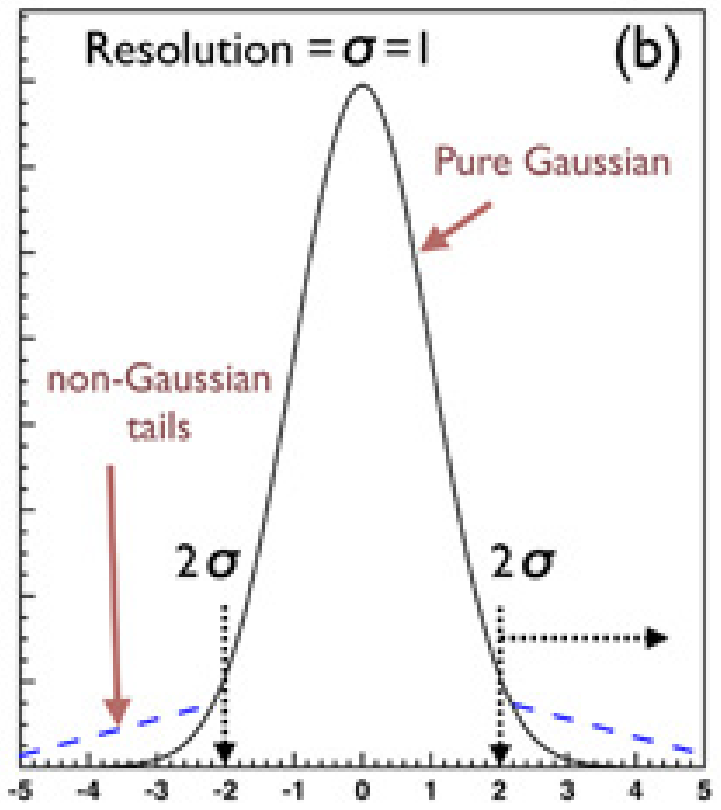}}    
\caption{(a) Time-of-flight differences for pairs of particles
plotted against momentum; (b) Illustration of a Gaussian resolution
function and example of non-Gaussian tails. \label{fg:tof}}
\end{figure}

\begin{figure}[ht] 
\centerline{\epsfxsize=4.1in\epsfbox{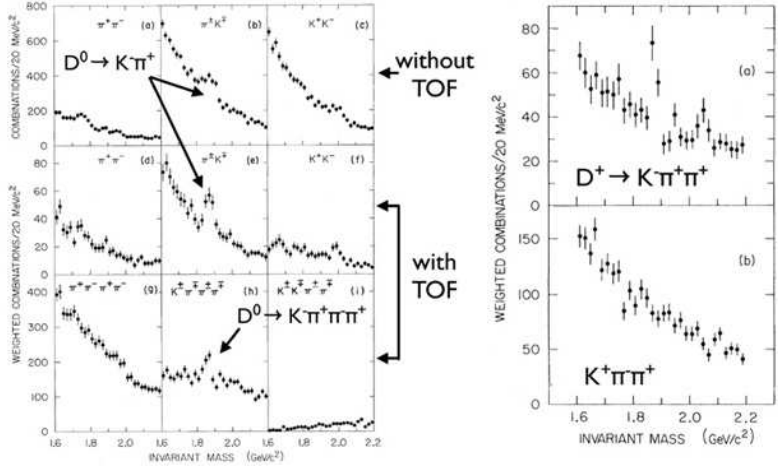}}    
\caption{Mass plots from Mark I:
(Left) $D^0$; (right) $D^+$. \label{fg:opencharm_plots}}
\end{figure}

Figure \ref{fg:tof}(a) shows the difference in time-of-flight over one
metre for pairs of hadrons. The performance of a {\em TOF} system is given
by the distance ($L$) traveled between the two time measurements and 
the resolution ($\sigma_{\Delta t}$) with which the
time-of-flight measurement is made. Long distances and fine resolution
are needed. For example for Mark I $L\approx 2$~m and
$\sigma_{\Delta t}\approx 400$~ps. This means that one can get
$2\sigma_{\Delta t}$ separation between kaons and pions for
momenta $<1$~GeV/c, {\em i.e.}
at very low momentum. Even if the time measurement
resolution can be considerably reduced, {\em e.g.} to $\approx$100~ps for
the Fermilab
CDF Run II experiment, it can be seen from Fig.~\ref{fg:tof}(a) that
with $L\approx 2$~m, a 
$2\sigma_{\Delta t}$ separation between kaons and pions
is only achieved for momenta $<2$~GeV/c. Even with only a
$2\sigma_{\Delta t}$ separation at low momentum, the decays
$D^0\rightarrow K^-\pi^+$, $D^0\rightarrow K^-\pi^+\pi^+\pi^-$ and
$D^+\rightarrow K^-\pi^+\pi^+$ could be isolated 
sufficiently from background
at Mark I for them to make the
discovery\rlap.\,\cite{ref:mark1opencharm}
The mass plots are shown in Fig.~\ref{fg:opencharm_plots}, the
$D^0\rightarrow K^-\pi^+$ distribution is shown without and with a
{\em TOF} kaon selection.

The gas {\em \v Cerenkov counters} mentioned in Part I can be used
to separate kaons and pions at higher momenta, but typically
collider experiments like Mark I and CDF do not have 
the necessary space for them. There are a number of alternate
particle identification systems\rlap.\,\cite{ref:textbooks}

\subsection{Measurement Uncertainties}

At this point it is worth making an aside about experimental
resolution and the meaning of a $2\sigma_{\Delta t}$ separation.
The importance of mass resolution was introduced in 
Sec.~\ref{sec:resolutions1}. Not only is the size of 
the resolution important, but the resolution function or shape matters also.
When a quantity is measured experimentally
one does not typically obtain an exact number, but there is some
uncertainty. This uncertainty is normally separated into two components.
One component is essentially statistical in nature and arise due to a lack
of precision. The other is typically non-statistical and is due to our
limited knowledge and affects the accuracy of the measurement. The former is
called the statistical uncertainty or statistical error, while the latter
is referred to as the systematic uncertainty.

A classic example of a statistical uncertainty is that
due to limited statistics.
{\em E.g.} when measuring the lifetime of a particle we have a limited
number of particles to use, thus the lifetime distribution is measured
with limited precision which leads to an uncertainty in the extracted
lifetime. Another example is measuring a distance with a measuring tape.
There is some uncertainty in positioning the tape at one end and 
in the reading and the precision of the scale at the other end. To reduce
this uncertainty the measurement can be repeated many times and the
average value used. We can illustrate the resolution with this simple
example. If a histogram is made of these measurements (frequency {\em vs.}
[value$-$nominal]) ideally the distribution is Gaussian as shown by the
solid line in Fig.~\ref{fg:tof}(b). The resolution is the sigma of the
Gaussian distribution, and it gives the 
statistical uncertainty of any single measurement.

Imagine Fig.~\ref{fg:tof}(b) shows the distribution of time-of-flight
measurements for pions, where the nominal value is subtracted off and
normalized to the resolution ($(\Delta t-\Delta t_0)/\sigma_{\Delta t}$).
For any given pion the $\Delta t$ measured can fall anywhere within the
Gaussian distribution. In particular for a small fraction of the time
it could be larger than 2$\sigma_{\Delta t}$ from nominal, for a Gaussian
distribution this probability is about 3\%. Suppose that for a kaon
the measured value of
$(\Delta t-\Delta t_0)$ is greater than 2$\sigma_{\Delta t}$. Then by
requiring $(\Delta t-\Delta t_0)> 2\sigma_{\Delta t}$ we can select
kaons and reject 97\%\ of pions. This is for an ideal Gaussian
distribution. The resolution function 
typically has non-Gaussian tails that go out much further as
illustrated 
crudely by the dashed lines in Fig.~\ref{fg:tof}(b). The rejection in
this case would not be as good as 97\%. Thus one needs to know the 
resolution function and must take care to try to avoid large
non-Gaussian tails. For a real {\em TOF} system, non-Gaussian tails
could arise from a number of sources, and the tails could also be
asymmetric. Some of these sources include the following.

\begin{itemize}
\item The counter giving the time signal is finite in size and the measurement
will depend on where the particle hits the counter.
\item The system is made up of many counters whose relative timing 
and locations are not perfect.
\item The calibration is not perfect, {\em e.g.} calibration tracks
do not always come from exactly the same point, and the start time is
not perfectly known.
\item Some effects like MCS may affect the resolution and cause it to
vary with the particle momentum.
\end{itemize}

Further coverage of statistical uncertainties and how to determine and
handle them are beyond the scope of this mini-course, but there are many
excellent books on this subject\rlap.\,\cite{ref:statbooks}

The other component of a measurement uncertainty is called the
systematic uncertainty and it is typically not statistical in
nature. A classic example can again be illustrated by the case of
measuring a distance with a measuring tape. If the scale of the
measuring tape is wrong we would get a systematic error. Of course,
if it were known that the scale was incorrect, we would correct the
scale and the systematic error would be eliminated. Now let us assume
that we must calibrate the measuring tape ourselves. We can only
calibrate the scale within a certain accuracy, and this leads to
a systematic uncertainty in the distance measured. 
For a more realistic
example consider measuring the lifetime of a decaying particle.
For a short lived particle like the $D^0$, the time is not directly
measured. Instead, the distance ($L$) traveled between production and
decay is measured and the momentum of the particle is also measured.
The proper time for the decay is then $t=Lm_{D^0}/p_{D^0}$.
Besides the length and momentum scales, there are other potential
sources of systematic uncertainties. The lifetime is extracted from
a lifetime distribution containing many particle decays. This distribution
may not be a pure exponential but could be modified due to
detector acceptance and efficiency. The correction for acceptance
and efficiency is typically determined using a Monte Carlo simulation.
There are inherent uncertainties in the simulation
that lead to an uncertainty in the
correction
function and thus to a systematic uncertainty in the lifetime. If the particle
passes through matter before decaying
or the daughter particles pass through matter, 
the lifetime distribution can also be affected by absorption of the
parent or daughter particles. The cross sections for absorption may be
poorly measured or not even known. This limited knowledge can also
lead to a systematic uncertainty in the measurement. Finally, another
source of systematic uncertainty
could be backgrounds that mimic the signal but which are not
properly accounted for. Typically, systematic uncertainties are not
well defined and are not straightforward to determine. They are also
usually not
Gaussian distributed, and combining systematic
uncertainties from different sources is problematic. Since even the
meaning and definition of systematic uncertainties are difficult to quantify,
ideally one should
design an experiment to have a small systematic uncertainty (compared
to the statistical uncertainty), so as not to have to worry about
the details of the treatment and combining of systematic uncertainties.
Further coverage of systematic uncertainties is beyond the scope of
this mini-course. The understanding of systematics is beginning to be
better understood and in some rare cases are even correctly 
taught\rlap.\,\cite{ref:systematicsarticle}
However considerable disagreements are still common.

\subsection{Improving S/B for Open Charm}

Although the use of particle identification can be powerful in isolating a
signal, it can be seen from Fig.~\ref{fg:opencharm_plots} that 
there is considerable room for improvement. This is especially true in
hadronic interactions which typically have higher backgrounds than in
$e^+e^-$ annihilations.

The lifetimes of the open charm particles are in the range
0.1--1~ps, which is small but finite and can be used to isolate a signal.
Almost all the $u$- and $d$-quark backgrounds have essentially zero 
lifetime while the backgrounds
from some strange particles decay after a long distance. 
Thus the signature of a charm
particle is given by its decay a short distance away from the production
point. For example, a 30~GeV $D^0$ travels an average length of about 2~mm,
which is quite small but increases linearly with momentum.

\begin{figure}[ht] 
\centerline{\epsfxsize=4.1in\epsfbox{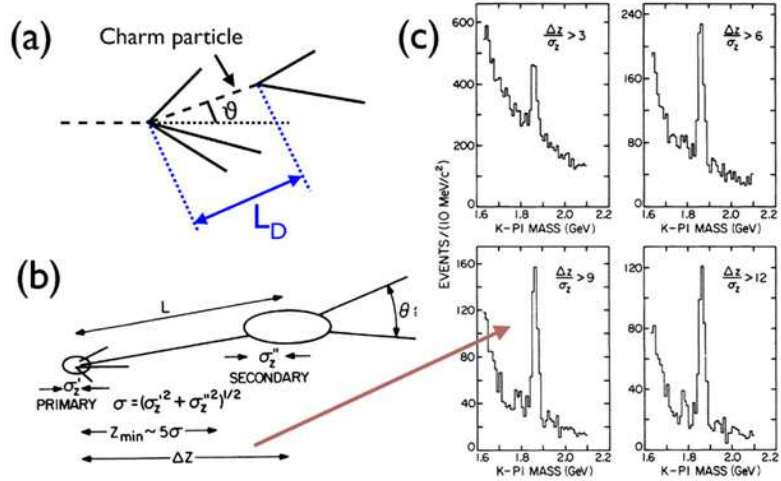}}    
\caption{(a),(b) Illustration of production and decay of a charm particle.
(c) Invariant $K^-\pi^+$ mass plots from E691 showing the power of a
detached vertex requirement. \label{fg:charmdecay_vertex}}
\end{figure}

To get a better sense of the scale involved, consider the decay of a
a charm particle produced in a fixed-target experiment 
as illustrated in Fig.~\ref{fg:charmdecay_vertex}(a).
The charm particle is produced and then decays
after traveling a distance $L_D$. To separate the production and decay
vertices we need to measure $L_D$ with a resolution of $\sigma_{L_D}<< L_D$.
Since position detectors typically measure in the dimension transverse
to the beam direction, it is more convenient to transform this
essentially longitudinal resolution requirement into an transverse one.
The typical angle that the charm particle is produced relative to the
beam direction is $\theta\approx m_D/p_D$, where $m_D$ and $p_D$ are the
mass and momentum of the charm particle respectively. The mean distance
traveled by the charm particle is $L_D=\beta\gamma c\tau_D=c\tau_Dp_D/m_D$
where $\tau_D$ is the lifetime of the charm particle. Thus to resolve the
production and decay vertices we need $\sigma_{\mathrm trans}<<\theta L_D$,
or $\sigma_{\mathrm trans}<<c\tau_D$, where $\sigma_{\mathrm trans}$ is the
transverse position resolution of the detector (charged) tracking system.
The values of $c\tau_D$ for the
$D^0$, $D^+$ and $\Lambda_c^+$ are 123~$\mu$m, 312~$\mu$m and 60~$\mu$m
respectively.

The resolution of the {\em MWPC's}
depend on the wire spacing ($s$), and for a single detector plane is 
given by $\sigma_{\mathrm trans}=s/\sqrt{12}$.
The minimum wire spacings are in the range
1--2~mm depending on their cross sectional coverage. For $s=2$~mm,
$\sigma_{\mathrm trans}=577$~$\mu$m, too large to resolve
the production and decay vertices.
The spatial resolution can be improved by measuring
the time between a charged particle passing through the detector plane
and when a signal is received in the wire closest to the point of passage.
This is done in {\em Drift Chambers} and resolutions as low as
$\sigma_{\mathrm trans}\approx 100$~$\mu$m have been obtained in such a
tracking system. This is still too large, especially considering that one
typically needs better than 5--10$\sigma$ vertex separation.

A tracking system with much better spatial resolution is needed. 
Historically, two 
detector technologies have been used that can give better
resolutions: {\em photographic emulsions} and 
{\em Bubble Chambers}.

Detection using layers of photographic emulsions has been used for 
a long time and spatial resolutions of better than 10~$\mu$m have been
obtained. Although these have been used relatively recently in 
DONUT to make the first direct observation of the 
$\nu_{\tau}$\cite{ref:donut}
they are not suitable for high rates.

{\em Bubble Chambers} have also been used historically to make important
observations. Typically, the resolution of bubble chambers
is not better than that for {\em Drift
Chambers}. However, a sufficiently small bubble chamber, like the LEBC in
the LEBC-EHS experiment, has achieved resolutions of $\sim$10~$\mu$m, but
again such bubble chambers
are not suitable for high rates. The LEBC-EHS experiment
reconstructed about 300-500 charm decays\rlap.\,\cite{ref:lebc}

What launched the high statistics studies of charm
quark physics was the development and use of the {\em Silicon Microstrip
Detector} ({\em SMD}). The Fermilab E691
photoproduction
experiment included one of the first
{\em SMD's} and collected
a 10,000 sample of fully reconstructed charm decays,
about two orders of magnitude more than other experiments of that time.
Resolutions as good as 
$\sigma_{\mathrm trans}\sim 10$~$\mu$m can be obtained and some
data from E691 are shown in 
Fig.~\ref{fg:charmdecay_vertex}(c)\rlap.\,\cite{ref:e691}
{\em SMD's} have now been used in many experiments including those studying
bottom and top quarks.

\subsection{Going for Higher Statistics}

\begin{figure}[ht] 
\centerline{\epsfig{file=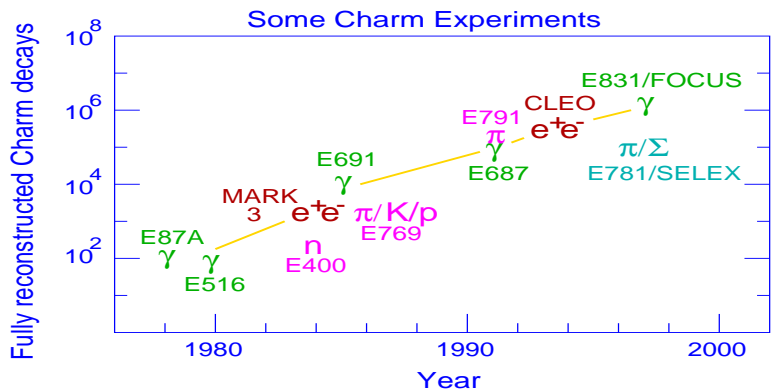,width=4.0in,height=2.0in}}
\caption{Number of fully reconstructed charm decays for different
experiments as a function of time. \label{fg:charm_road}}
\end{figure}

The road to higher statistics in charm studies is illustrated in 
Fig.~\ref{fg:charm_road}, giving some selected milestones
along the route.

The charm quark experiments include $e^+e^-$ and $p\bar{p}$ colliders
as well as fixed-target experiments using photon and hadron beams.
While larger charm data sets could be obtained in  $e^+e^-$ experiments
by increasing the luminosity ({\em e.g.} CLEO),
the fixed-target experiments needed additional technological advances --
the 8~mm tape for data storage and high power commodity computing for data
processing. For example, using these technologies and by building a more
intense photon beam,
the Fermilab FOCUS photoproduction experiment obtained
a sample of 1 million fully reconstructed charm decays with published
physics results one year after the end of data taking.
Using 8~mm tapes with 30 times the capacity of 9-track tapes, the
Fermilab E791 hadroproduction experiment could write out much more
data and collected more charm than previous hadroproduction experiments.
To do this,
E791 used a wall of 42 8~mm tape drives in parallel to record
data fast enough.
To obtain substantially more statistics, a revolution in triggering is
needed.
 
\subsection{The Trigger System and the Bottom Quark}

Typically the particle interactions occur at a high rate and the S/B can be
as low as 10$^{-3}$--10$^{-8}$. A trigger system is used to quickly decide
whether an interaction contains signal and thus 
``trigger'' the recording of the related data.

For a charm photoproduction experiment like FOCUS,
the photon beam largely produces $e^+e^-$ conversion pairs and only
about 1 in every 500 photon interactions
would produce hadrons. Only 1 in every 150
of those interactions producing hadrons contains a charm quark.
It is relatively
easy to recognize a photon conversion from an interaction producing hadrons.
Since the fraction of hadron producing interactions containing charm 
is not too small, one just writes out all hadron producing interactions.
For hadroproduction experiments on-the-other-hand, the fraction of charm
is smaller by another factor of ten, thus either a lot more data must be
written out and analyzed or a better, more intelligent trigger must be used.

Historically 
the trigger is a system of fast electronics that quickly processes special
trigger signals produced by the detector and gives an
electronic acceptance decision that is used to ``trigger'' the readout
to save the data for that interaction. Long cables are used to delay the
signals from the rest of the detector so they do not arrive at the
readout before the trigger decision is made. In addition, the experiment
is ``dead'' and unavailable to collect more data until the data readout
is completed -- this dead-time can be a significant fraction of the
live-time.
In the first stage of
the FOCUS trigger, the decision must be made within about 370~ns from
the time the interaction occurs.

Developments have made modern trigger systems much easier. Electronics
are now faster, smaller and cheaper. Also, high speed data links and
computing resources are more
powerful. A large amount
of memory is now affordable so that data from the detectors can be stored
digitally while the trigger processing takes place. This eliminates
the need for long signal cables which can degrade analog signals, and it also
gives more time for trigger processing and virtually eliminates
dead-time.

The ultimate trigger is if all the data could be 
recorded and analyzed before deciding
which data to store. One can illustrate what is needed for such a trigger
by using as an example the CDF or D0 experiments
at Fermilab. In these experiments,
protons and antiprotons cross every 396~ns, so there
are about
2.5$\times$10$^6$ crossing/s. If it took one second to fully analyze the
data from one crossing in a single CPU, we would need 2.5 million
CPU's to not lose data from any crossings. We would also need to
temporarily store at least 2.5 million crossings worth of data.
If one needs 300~KB/crossing, then 1~TB (10$^3$~GB) is needed. Since
the processing time would have a long tail beyond one second, to be safe
one would want about 1000~TB of RAM as well as the 2.5 million CPU's!
Clearly a trigger that only partially processes the data is needed.

Even if the ultimate trigger cannot yet be realized,
the developments mentioned above have provided the needed ingredients
to separate out charm decays in hadronic collisions by looking
for evidence of a detached decay vertex at the trigger level. Since 
recognizing bottom quark decays is similar to that for charm decays,
this revolution in triggering has made possible an experiment
that will reconstruct very large samples of charm and bottom decays.

Already large samples of charm and bottom quark decays are being
collected using the BaBar and Belle $e^+e^-$ experiments. To do better
one must use hadronic collisions with sufficient energy like
$p\bar{p}$ annihilations at the Fermilab Tevatron. The cross section
for producing bottom quarks is much larger than in $e^+e^-$
annihilations, {\em e.g.} about 100~$\mu$b compared to about 1.1~nb
at the $\Upsilon$(4S). The CDF and D0 $p\bar{p}$ experiments can
collect sizable samples of charm and bottom decays, but to get 1000
times more rate than BaBar or Belle requires a specialized detector and
data acquisition and trigger systems.

The BTeV experiment\cite{ref:btev} is designed to study bottom and
charm decays at the Tevatron. 
To maximize the yield for clean flavour-tagged $B$ mesons 
for CP violation studies, the detector is
placed in the forward direction allowing 
a {\em Ring Imaging \v Cerenkov Counter} ({\em RICH}) for
excellent particle identification
over a wide momentum range. A PbWO$_4$ crystal calorimeter provides
efficient detection of photons and $\pi^0$'s with excellent energy
resolution. The BTeV experiment includes a {\em silicon pixel
detector} that makes possible the recognition of detached vertices at
the lowest trigger level.

\begin{figure}[ht] 
\centerline{\epsfxsize=3.5in\epsfbox{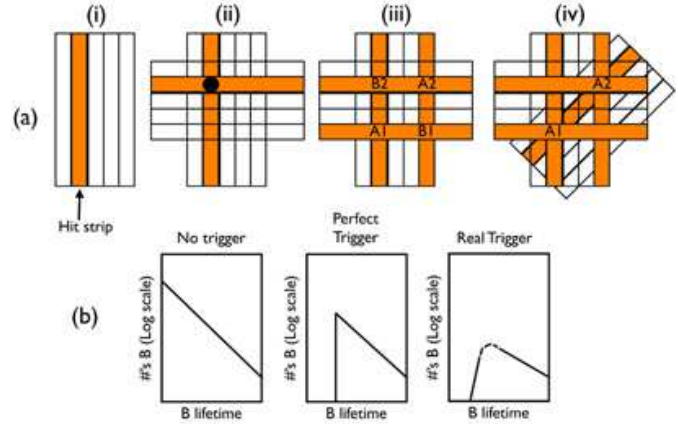}}    
\caption{(a) Illustration of pattern recognition in a {\em SMD}.
(b) Illustration of the effect on the lifetime
distribution of a detached vertex trigger. \label{fg:trigger_illustration}}
\end{figure}

Although a {\em SMD} can provide excellent spatial resolution, a lot of
data processing is typically needed for interactions with many tracks
due to the strip geometry. This is illustrated in 
Fig.~\ref{fg:trigger_illustration}(a). A single particle passing through
a plane of strips will give a signal in one strip as illustrated in (i).
The location along the hit strip can be determined by a second plane of strips
oriented at 90$^{\circ}$ to the first plane as in (ii). However if,
as illustrated in (iii), two
particles pass through the two planes we would get four hit strips and one
cannot tell if the two 
particles passed through points (A1, A2) or through (B1, B2). This
ambiguity may be resolved by a third plane of strips at an angle as
given in (iv). For a complex event with many particles the pattern 
recognition becomes quite complex and requires significant CPU power.

Ideally a trigger algorithm should be
close to that used in the data analysis
but with looser selection criteria. This is because a
poorly chosen trigger algorithm can give rise to sizable systematic
uncertainties. A simple example is illustrated in 
Fig.~\ref{fg:trigger_illustration}(b). The lifetime distribution of
the $B^0$ meson is shown which is a pure exponential with the $B^0$ lifetime.
Since backgrounds are typically at low lifetimes, an ideal trigger for
collecting data to measure the lifetime would select decays with
a large enough lifetime as illustrated in the middle distribution
of Fig.~\ref{fg:trigger_illustration}(b). In a real trigger there
is typically only time to do partial processing, and thus one
might require only the
presence of one or two detached tracks, instead of reconstructing the
production and $B^0$ decay vertices.
This could lead to a lifetime
distribution illustrated by the right-most distribution of
Fig.~\ref{fg:trigger_illustration}(b). Thus a correction function is
needed to extract the correct $B^0$ lifetime.

The charged track pattern recognition is simplified 
in BTeV by the use of 400$\times$50~$\mu$m$^2$
{\em silicon
pixels} which can locate the position of a passing particle 
in 3-dimensions by a single hit pixel. 
Low momentum tracks undergo more MCS and
can give rise to false detached tracks. 
In BTeV, the pixel detector is located in a
dipole magnet so that tracks with low momentum can be rejected and not
used at the trigger level. 

Even with 
{\em silicon pixels} a full reconstruction cannot be done. 
Custom electronics
using 500 FPGA's are used to help in processing the 500~GB/s data
rate coming from the detectors. Further
data processing and the
pattern recognition is done on 500 commercial IBM-G5-equivalent 
processors.
Two further levels of the trigger running on 1500 commodity CPU's
reduce the data going to storage to a more manageable 200~MB/s. 

The BTeV experiment nicely illustrates the convergence of a number of
technological advances. Years of scientific progress
have enabled such an experiment to be
realized.


\section*{Glossary}

{\flushleft{\bf\em Bubble chamber:}}
A historic detector consisting of a liquid maintained at a
pressure above the equilibrium vapour pressure. The bubble chamber
can be expanded to suddenly decrease the pressure so that charged
particles passing through the liquid in a ``superheated'' condition
will create a track of bubbles. Photographs are taken of the bubbles in
multiple views to reconstruct the particle trajectories.

{\flushleft{\bf\em Calorimeter:}}
A device to measure the energy of particles. The two distinct types
are {\em electromagnetic calorimeters} and hadronic
calorimeters. They work by completely absorbing the
shower produced by a particle and
producing a signal proportional to its energy. 
Calorimeters must be calibrated
to give the absolute particle energy.

{\flushleft{\bf\em \v Cerenkov counter:}}
A detector based on the \v Cerenkov effect (for which \v Cerenkov shared
the 1958 Nobel prize). Particles traveling faster than light in a given
medium emits a cone of (\v Cerenkov) light. A threshold \v Cerenkov
counter contains a gas,
for example, with a well chosen refractive index so that
for a given particle momentum one particle type ({\em e.g.} pions) will
emit light while another ({\em e.g.} kaons) will not. The angle of the
cone of light also depends on the particle velocity which is used in
other forms of {\em \v Cerenkov counters} like the {\em RICH}.
The amount of \v Cerenkov light emitted is typically low, about
100 times less intense
than scintillation light in a scintillator.

{\flushleft{\bf\em Drift chamber:}}
A {\em wire chamber} where one measures the
time between when a charged particle passes
through and when a signal in the nearest signal wire is received.
Typically many wires are used to form drift cells where the
electric field is tailored to obtain a fairly uniform drift velocity
across the cell. The spatial resolution is better than a {\em MWPC}
but a drift chamber is more complex and typically cannot handle as
high a rate of particles.

{\flushleft{\bf\em Electromagnetic calorimeter:}}
A {\em calorimeter} for measuring the energies of photons and $e^{\pm}$
through their electromagnetic interactions. These calorimeters can be
made from dense crystals like
PbWO$_4$ or {\em lead-glass}, or can be sandwiches made 
of multiple layers of dense absorber and detection material.

{\flushleft{\bf\em Emulsions:}}
Usually a layer of photographic emulsion several hundred $\mu$m thick
in which a 
traversing charged particle causes the 
nearest silver halide grains to develop.
Each grain is typically 0.2~$\mu$m in diameter with about 270 developed
grains/mm. The emulsion must be scanned to reconstruct the particle
trajectories.

{\flushleft{\bf\em Lead-glass shower counters:}}
A dense glass used to detect photons and $e^{\pm}$ and for
{\em electromagnetic calorimeters}. The detection is based on
\v Cerenkov light.

{\flushleft{\bf\em Magnetic Spectrometer:}}
A detector system used to determine the momentum of charged particles
by measuring the defection of the particles in a known magnetic field.
Various magnetic field configurations can be used {\em e.g.} dipole,
solenoid, or toroid. Deflection of particles are measured using position
detectors, usually {\em wire chambers}, but can be {\em e.g.}
{\em scintillator hodoscopes}, {\em scintillating fibres}
or a {\em SMD}.

{\flushleft{\bf\em Multiwire proportional chamber (MWPC):}}
A {\em wire chamber} where the location of a passing charged particle
is determined by the location of the wire closest to it.

{\flushleft{\bf\em Photomultiplier Tube (PMT):}}
A device to detect a small quantity of light using the
photoelectric effect (for which Einstein received the 1921
Nobel prize). The maximum sensitivity of
the photocathode in a typical {\em PMT} is for blue light.

{\flushleft{\bf\em Ring imaging \v Cerenkov counter (RICH):}}
A {\em \v Cerenkov counter} where the angle 
of the emitted \v Cerenkov light is measured
to enable the identification of particles over a wide momentum range.

{\flushleft{\bf\em Scintillator:}}
A material that produces light through fluorescence when a charged particle
passes through it. Scintillators used include inorganic crystals like
PbWO$_4$, organic liquids and plastic. A classic plastic scintillator
is made of polystyrene that produces light in the UV. The UV light is
shifted to blue with a tiny doping of primary and secondary fluors
to better match the photosensitivity of a {\em PMT}.
Most of the light comes in a fast component (few ns) and strong signals
are possible with sufficient scintillator thickness.

{\flushleft{\bf\em Scintillator fibres:}}
{\em Scintillator} in the form of long flexible fibres with an outer acrylic 
sleeve so that the scintillation light is 
isolated to the fibre, but still
totally internally reflected
along the fibre to the ends. Typically a few mm in diameter they are used
for position detectors or in {\em calorimeters} as either the detection
material or as a mechanism for readout.

{\flushleft{\bf\em Scintillator hodoscope:}}
A single detector plane made of strips of 
{\em scintillator}. Used to detect the
position of a charge particle. Two planes
can be overlapped 
with the strips in one plane oriented at 90$^{\circ}$ to the other
to locate the particle in both transverse dimensions.

{\flushleft{\bf\em Silicon microstrip detector (SMD):}}
Detection is based on essentially a silicon semiconductor p-n
junction where the depletion region is enlarged by a bias voltage.
The depletion layer can be considered as a solid state ionization chamber.
A charged particle passing through the depletion region
liberates electron-hole pairs which 
create signals on very thin, closely
spaced readout strips. {\em SMD's} have the detection regions arranged as long 
uniformly separated strips. The strip separation can be in
the range 10--300~$\mu$m.

{\flushleft{\bf\em Silicon pixels:}}
Similar to the {\em SMD} but the active region is in the form of
rectangles so that a ``hit'' pixel locates a passing particle in
both transverse dimensions. The readout is however more complicated.

{\flushleft{\bf\em Spark wire chamber:}}
A parallel-plate gas chamber
in which a high voltage pulse is applied immediately after the passage of
a passing charged particle. Sparks form along the trail of 
ions caused by the charged particle passing through the gas. 
This can provide a visualization of
the track useful for public demonstration. High speed readout is
typically done magnetostrictively or capacitively.

{\flushleft{\bf\em Time-of-flight detector:}}
A system for identifying charged particles based on measuring their velocity
between two points. The time-of-flight between two points is usually
measured using {\em scintillator} counters possibly in conjunction with a
measurement of the time of an interaction. The particle
momentum is also measured giving a velocity that can distinguish particle
types through their differing masses. See Sec.~\ref{sec:opencharm}.

{\flushleft{\bf\em Wire chamber:}}
For detection of charged particles through their ionization of
usually noble gas atoms. A high voltage causes ionized electrons to
accelerate and create an avalanche of electrons and positive ions.
Detection of the avalanches in a plane of wires can give the position
of the passing particle.
Many types of wire chambers have been used, the original type ({\em MWPC})
was invented by Charpak
for which he received the 1992 Nobel prize.

\section*{Acknowledgments}
My thanks to Jeff Appel for some helpful
suggestions for these lectures and for a careful reading of this writeup. 
This work was supported by the Universities Research Association
Inc. under Contract No. DE-AC02-76CH03000 with the U.~S.
Department of Energy.


\begin{thebibliography}{99}
\bibitem{ref:textbooks} Some examples of textbooks are:
R.~Fernow, {\it Introduction to experimental particle
physics}, CUP 1986; K.~Kleinknecht, {\it Detectors for particle radiation},
2nd Ed., CUP 1998. There are also some excellent articles, {\em e.g.} in
F.~Sauli (Ed.), {\it Instrumentation in High Energy Physics},
World Scientific, 1992.

\bibitem{ref:charmdiscovery}R.~N.~Cahn and G.~Goldhaber, {\it The experimental
foundations of particle physics}, CUP 1989.

\bibitem{ref:tingnobel} S.~C.~C.~Ting, Nobel Lecture, 11 Dec. 1976, the
full text is available at
\url{http://nobelprize.org/physics/laureates/1976/ting-lecture.pdf}.


\bibitem{ref:richternobel} B.~Richter, Nobel Lecture, 11 Dec. 1976, the
full text is available at
\url{http://nobelprize.org/physics/laureates/1976/richter-lecture.pdf}.

\bibitem{ref:lederman1970}J.~H.Christenson {\em et al.}, 
{\it Phys. Rev. Lett.} {\bf 21}, 1523 (1970).

\bibitem{ref:ledermannobel} L.~M~Lederman, Nobel Lecture, 8 Dec. 1988,
the full text is available at
\url{http://nobelprize.org/physics/laureates/1988/lederman-lecture.pdf}.

\bibitem{ref:tingpapers} J.~J.~Aubert {\em et al.}, 
{\it Phys. Rev. Lett.} {\bf 33}, 1404 (1974); 
{\em Nucl. Phys.} {\bf B89}, 1 (1975).

\bibitem{ref:e835} M.~Ambrogiani {\em et al.},
{\it Phys. Rev.} {\bf D64}, 052003 (2001).

\bibitem{ref:lederman1976} D.~C. Horn {\em et al.},
{\it Phys. Rev. Lett.} {\bf 36}, 1236 (1976).

\bibitem{ref:lederman1977} S.~W.~Herb {\em et al.},
{\it Phys. Rev. Lett.} {\bf 39}, 252 (1977).

\bibitem{ref:ledermanneutrino} G.~Danby {\em et al.}, 
{\it Phys. Rev. Lett.} {\bf 9}, 36 (1962). See also the Nobel lectures at
\url{http://nobelprize.org/physics/laureates/1988/}

\bibitem{ref:mark1opencharm} G.~Goldhaber {\em et al.},
{\it Phys. Rev. Lett.} {\bf 37}, 255 (1976);
I.~Peruzzi {\em et al.}, {\it Phys. Rev. Lett.} {\bf 37}, 569 (1976).

\bibitem{ref:statbooks} Some examples of statistics textbooks
are: P.~R.~Bevington
and D.~K.~Robinson, ``Data
Reduction and Error Analysis for the Physical Sciences'', 2nd Ed.,
McGraw-Hill 1992;
G.~Cowan, ``Statistical Data Analysis'', OUP 1998;
D.~S.~Sivia, ``Data Analysis: A Bayesian Tutorial'', OUP 1996.

\bibitem{ref:systematicsarticle} R.~Barlow, ``Systematic Errors: Facts
and Fictions'', hep-ex/0207026.

\bibitem{ref:donut} K.~Kodama {\it et al.},
{\it Phys. Lett.} {\bf B504}, 218 (2001).

\bibitem{ref:lebc} M.~Aguilar-Benitez {\em et al.}, 
{\it Z. Phys.} {\bf C40}, 321 (1988).

\bibitem{ref:e691} J.~C.~Anjos {\em et al.}, {\it Phys. Rev. Lett.} 
{\bf 58}, 311 (1987); K.~Sliwa {\em et al.}, {\it Phys. Rev.}
{\bf D 32}, 1053 (1985); J.~R.~Rabb {\em et al.} {\it Phys. Rev.}
{\bf D 37}, 2391 (1988).

\bibitem{ref:btev} \url{http://www-btev.fnal.gov/}

\end{thebibliography}
\end{document}